\documentclass[dvips,twoside,fleqn]{article}
\usepackage{times}
\usepackage{amstext}
\usepackage{amssymb}
\usepackage{espcrc2}
\usepackage{graphicx}
\title{Topology in QCD with 4 flavours of dynamical fermions
\thanks{Partially supported by MURST and by EC TMR program ERBFMRX-CT97-0122.}
}

\author{B.~All\'es\address{Dipartimento di Fisica, 
Universit\`a di Milano--Bicocca and
INFN--Milano, I--20133 Milano, Italy},
M.~D'Elia\thanks{Speaker at the conference.}\address{Dipartimento di Fisica, Universit\`a
di Pisa and INFN--Pisa, I--56127 Pisa, Italy},
A. Di Giacomo$^{\rm b}$.}

\begin{document}
\pagestyle{empty}

\begin{abstract}
We study the topological properties of full QCD with four flavours of dynamical
staggered fermions. In particular the topological susceptibility
is measured and the problem of the determination of its first derivative
is discussed.
\end{abstract}

\maketitle

\section{INTRODUCTION}

Topology plays an essential role in determining
the low energy features 
of strong interaction physics.
In this paper we present a lattice study of the topological 
properties of QCD in presence of four
flavours of degenerate dynamical staggered fermions.

A relevant quantity is the
topological susceptibility, defined in the continuum as
\begin{equation}
 \chi \equiv \int d^4x \;\; \partial_\mu 
   \langle 0 | {\rm T}\left\{K_\mu(x) Q(0)\right\}| 0 \rangle \; ,
\label{eq:chi}
\end{equation}
where $K_\mu(x)$ is the Chern current
and $Q(x)=\partial_\mu K_\mu(x)$ is the topological charge density,
\begin{equation}
 Q(x) = {g^2\over 64\pi^2} \; \epsilon^{\mu\nu\rho\sigma} \;
        F^a_{\mu\nu}(x) \; F^a_{\rho\sigma}(x)  \; .
\end{equation}
Eq.~(\ref{eq:chi}) defines the prescription for the singularity 
of the time ordered product when $x \rightarrow 0$~\cite{witten}.

The value of $\chi$ in the quenched theory is related 
to the $\eta '$ mass by the Witten-Veneziano formula. It has been
successfully measured on the lattice~\cite{tsu3}, confirming
the Witten-Veneziano prediction.

In full QCD with spontaneous 
chiral symmetry breaking, 
$\chi$ is related to the quark condensate~\cite{chiral}
\begin{equation}
\chi = {m_q \over N_f} \langle \bar{\psi} \psi \rangle_{m_q = 0} + o(m_q) \; ,
\end{equation}
where $m_q$ is the quark mass and $N_f$ the number of flavours. We want 
to test Eq. (3) for $N_f = 4$ and different values of $m_q$.
  
Another relevant quantity is the slope at $q^2 = 0$ of the 
topological susceptibility, $\chi '$, defined as
\begin{equation}
\chi' \equiv {d \chi(q^2) \over d q^2}\big|_{q^2 = 0} = {1 \over 8}
\int d^4x \langle Q(x) Q(0) \rangle x^2\;,
\end{equation}
where $\chi(q^2) = \int d^4x \; e^{i q \cdot x} 
   \langle Q(x) Q(0) \rangle $. In the quenched theory 
the consistency  of the Witten-Veneziano mechanism
requires $\chi'$ to be small. Instead in full QCD it is 
expected to be larger and related to the singlet axial charge
(and so to the proton spin crisis)~\cite{naris}. Therefore
its determination is of particular importance.
Unfortunately, techniques which
work well for the lattice determination of $\chi$, are not straightforwardly
applicable to the measurement of $\chi'$. We will present a preliminary
estimate of $\chi'$ and discuss perspectives for future refinements.

\section{TOPOLOGICAL SUSCEPTIBILITY}

We have simulated the full theory using the HMC algorithm
with 4 flavours of staggered fermions at $\beta = 5.35$ and 
four different values of the bare quark mass, $a \cdot m_q = 0.010,0.015,0.020,0.050$,
performing for each mass value respectively 6000,1500,1000 and 3000 units
of molecular dynamics time.
We have used a $16^3 \times 24$ lattice  and the Wilson action
for the pure gauge sector.

\begin{figure}[!ht]
\vspace{3.8cm}
\includegraphics{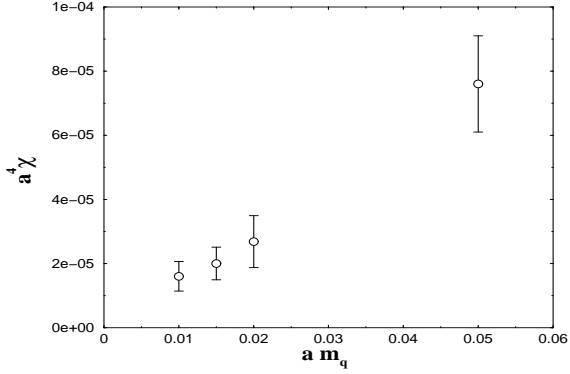}
\null\vskip 0.3cm
\caption{Topological susceptibility in lattice spacing units
at different bare quark masses and $\beta = 5.35$.}
\end{figure}

The field theoretical method has been used to determine $\chi$.
Given a discretization $Q_L(x)$ of the topological charge density, 
we define the lattice  susceptibility:
\begin{equation}
\chi_L = \sum_x \langle Q_L(x) Q_L(0) \rangle = \frac{ \langle Q_L^2 \rangle}{V}\;.
\end{equation}
$Q_L$ is related by a finite multiplicative
renormalization $Z$ to the continuum topological 
charge\footnote{In the full theory mixings of $Q_L$ with
fermionic operators appears,which anyway can be shown to be 
negligible~\cite{panag}.}. Moreover $\chi_L$ in general does not meet 
the continuum prescription used in Eq. (1) and, due to contact terms
in the product $Q(x)Q(0)$ at small $x$, a further additive
renormalization appears 
\begin{equation}
\chi_L = a^4 Z^2(\beta) \chi + M(\beta)\; .
\end{equation}
The heating method~\cite{heat} has been used to evaluate the 
renormalizations. $Z$ is determined by measuring $\langle Q_L \rangle$
on a sample of thermalized configurations
obtained by heating a semiclassical one-instanton configuration. 
Similarly $M$ is obtained by measuring $\chi_L$ on a sample with 
zero topological background, using the fact that Eq. (1) 
leads to $\chi = 0$ in the trivial topological sector.
Smearing techniques have been used to improve the operator $Q_L$~\cite{haris},
in order to reduce the renormalizations and obtain a better
accuracy on $\chi$ after subtraction. We have performed 
2 smearing steps, starting from
a standard discretization of $Q$, with a smearing
coefficient $c = 0.9$, as reported in ~\cite{tsu3}.

In Fig. 1 we report the results obtained for $a^4 \chi$ versus the quark
mass. The clear dependence of $a^4 \chi$ on $m_q$ 
disappears when going to physical units. We have fixed the scale
by measuring the string tension, 
obtaining $a^2 \sigma = 0.073(7)$ at $a \cdot m_q = 0.05$ and 
$a^2 \sigma = 0.033(4)$ at $a \cdot m_q = 0.01$. From these values, assuming
$\sqrt{\sigma} = 440 \;{\rm MeV}$, we obtain
$a(a \cdot m_q = 0.01) = 0.081(5) {\rm fm}$ and $a(a \cdot m_q = 0.05) = 0.0121(6) {\rm fm}$,
from which
$(\chi)^{1/4} (a \cdot m_q = 0.01) = (153 \pm 16) \;{\rm MeV}$ and 
$(\chi)^{1/4} (a \cdot m_q = 0.05) = (153 \pm 11) \;{\rm MeV}$. 
We cannot rule out possible systematic errors
coming both from the poor sampling of topological modes at 
the lowest quark masses~\cite{boyd} and from the determination of the physical
scale. Indeed at $a \cdot m_q = 0.01$ we have determined the lattice spacing
also by measuring  $m_\rho$ and $m_\pi$, obtaining a different
result, $a = 0.101(5) {\rm fm}$. Using this value 
we obtain $(\chi)^{1/4} (a \cdot m_q = 0.01) = (123 \pm 10) \;{\rm MeV}$,
which is closer to the theoretical expectation coming from Eq. (4),
$(\chi)^{1/4} \simeq 110 \;{\rm MeV}$.

\section{ESTIMATE OF $\chi'$}

On the lattice we can define
\begin{equation}
\chi_L' = {1\over8}\sum_x \langle Q_L(x) Q_L(0) \rangle x^2 \;.
\end{equation}
A relation similar to Eq. (6) holds in this case ~\cite{meggio}:
\begin{equation}
\chi_L' = a^4 Z^2(\beta) \chi' + M'(\beta).
\end{equation}
While $Z$ in the previous equation is the same appearing in Eq. (6),
since it is purely related to the renormalization of the topological
charge, $M'$ is a new additive renormalization, containing mixings of 
$\chi_L'$ to operators of equal 
or lower dimension. More generally, defining the two-point correlation
function of the topological charge density on the lattice, 
$\langle Q_L(x) Q_L(0) \rangle$, one can write
\begin{equation}
\langle Q_L(x) Q_L(0) \rangle = a^8 Z^2 \langle Q(x) Q(0) \rangle +
m(x) \; ; 
\end{equation}
$m(x)$ indicates mixings with terms of equal or
lower dimension in the Wilson OPE of $\langle Q_L(x) Q_L(0) \rangle$, 
and is related to $M$ and $M'$  in the following way:
\begin{equation}
M = \sum_x m(x); \;\;\; M' = \sum_x m(x) x^2\;
.
\end{equation}

\begin{figure}[!t]
\vspace{3.5cm}
\includegraphics{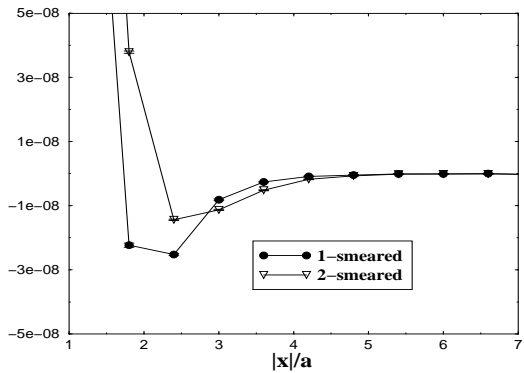}
\null\vskip 0.3cm
\caption{$\langle Q_L(x) Q_L(0) \rangle$ for the operators
after one and two smearing steps, at $a \cdot m_q = 0.01$}
\end{figure}

$M'$ cannot be determined by the heating method, since in this case
the continuum $\chi'$ is not constrained to be zero in 
the trivial topological sector. Therefore the techniques
used for $\chi$ cannot be straightforwardly applied to the determination of
$\chi'$. 
In principle $M'$ can also be computed by lattice
perturbation theory, but in practice this approach is not particularly
successful, especially when dealing with smeared operators for which
the convergence properties of the perturbation series worsen. 
We stress that also cooling techniques, which can usually
be used to determine $\chi$, fail in the determination of $\chi'$:
only the zero-moment of the two-point function
$\langle Q(x) Q(0) \rangle$, i.e. $\chi$, is topologically protected,
since it can be expressed in terms of the global topological charge,
$\chi = \langle Q^2 \rangle/V$, and $Q$ is quasi-stable under cooling. 
This is not the case for higher moments, and in particular for $\chi'$.

One possibility is to use an improved 
topological charge operator for which $M'$, even if not computable, 
is known to be small and thus negligible in Eq. (8). If we follow
this ansatz for the 2-smeared operator used previously we obtain, 
from our simulation at $a \cdot m_q = 0.01$, 
$\sqrt{|\chi'|} \sim 20 \;{\rm MeV}$,
in good agreement with the value expected from sum rules ~\cite{naris},
$\sqrt{|\chi'|} = (25 \pm 3) \;{\rm MeV}$.
However, the systematic error deriving from neglecting $M'$, $M'/\chi'$, 
can be estimated to be of the same order of magnitude 
as $M/\chi \simeq 40\%$ at $a \cdot m_q = 0.01$, that is still quite large.

A better estimate of $\chi'$ requires deeper knowledge 
of the two-point correlation function.
In the continuum  $\langle Q(x) Q(0) \rangle$ is known to be negative,
by reflection positivity, for $x > 0$, and positive and singular for $x = 0$.
Similarly, when using a lattice action which preserves reflection positivity,
we expect $\langle Q_L(x) Q_L(0) \rangle < 0$ whenever the two operators
$Q_L(0)$ and $Q_L(x)$ do not overlap. This is clear in Fig. 2, where
a determination at $a \cdot m_q = 0.01$ of 
$\langle Q_L(x) Q_L(0) \rangle $ averaged over spherical shells of width
$\delta x = 0.6 a$ is reported for 1 and 2 smearings.

\begin{figure}[!t]
\vspace{3.5cm}
\includegraphics{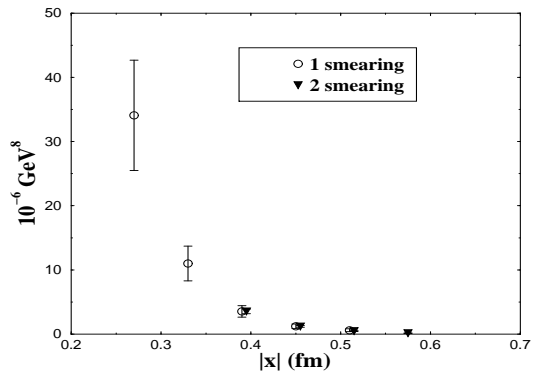}
\null\vskip 0.3cm
\caption{$ - \langle Q(x) Q(0) \rangle$ from 1
and 2-smeared operators at $a \cdot m_q = 0.01$}
\end{figure}

The information on $\langle Q_L(x) Q_L(0) \rangle $ is not enough
to extract  $\langle Q(x) Q(0) \rangle$: according to Eq. (9),
also $m(x)$ is needed. However we know that $m(x)$ comes from contact 
terms, so it must be zero at large $|x|$. Therefore, for a given
operator $Q_L(x)$, there must be an $x_0$ such that, for $|x| > x_0$,
$m(x)$ can be ignored in Eq. (9) and $\langle Q(x) Q(0) \rangle$ can be easily
extracted, the value of $Z$ being known from the determination of $\chi$.

A practical way to extract $x_0$ is the following: 
$\langle Q_L(x) Q_L(0) \rangle $ is determined for two operators
$Q_{L1}$ and $Q_{L2}$ and one looks for a plateau at large
$|x|$ in the ratio of the two functions, corresponding to the squared
ratio of the multiplicative renormalizations, $(Z_1/Z_2)^2$.   
In this way we have determined $\langle Q(x) Q(0) \rangle $ 
for 1 and 2 smearing steps at large $|x|$ and $a \cdot m_q = 0.01$, 
as shown in Fig. 3. There is good agreement between the two determinations,
as expected.
Work is in progress to extrapolate the information 
at large $|x|$ to smaller distances, 
thus allowing a more careful determination of $\chi'$.

\end{document}